\documentclass[12pt]{article}
\usepackage{graphicx}
\usepackage{amsmath}
\newcommand{\NN}{\hbox{I\kern-.2em\hbox{N}}}  %Naturali
\newcommand{\ZZ}{{{\rm Z}\kern-.28em{\rm Z}}} %Interi
\newcommand{\RR}{\mathop{{\rm I}\kern-.2em{\rm R}}\nolimits} %Reali
\newcommand{\QQ}{\hbox{l\kern-.36em\hbox{Q}}}  %Razionali
\newcommand{\CC}{\hbox{I\kern-.58em\hbox{C}}}

\begin{document}
\title{Locality of quantum mechanics and the
\\interpretation of the EPR criterion of reality.}
\author{Giuseppe Nistic\`o and Angela Sestito\\{\small
Istituto Nazionale di Fisica Nucleare, Italy}\\{\small Dipartimento
di Matematica, Universit\`a della Calabria, Italy}\\ \small{ email:
gnistico@unical.it, sestito@mat.unical.it}} \maketitle
 \abstract{We prove that by adopting a strict interpretation of the Einstein-Podolsky-Rosen
 criterion of reality, the proofs of the known non-locality theorems fail in
 showing that quantum mechanics violates the principle of locality and reality.}
%%%%%%%%%%%%%%%%%%%%%%%%%%%%%%%%%%%%%%%%%%%%%
\section{Introduction}
The different {\it non-locality} theorems appeared in the literature
\cite{1} - \cite{6}, starting from the pioneer theorem of Bell,
were, and are, often interpreted as proofs of inconsistency between
quantum mechanics and the {\it principle of locality and reality}
introduced by Einstein, Podolsky and Rosen \cite{7}, which consists
of the following two statements. \vskip.7pc\noindent (R) {\sl
Criterion of reality.}\quad{\it If, without in any way disturbing a
system, we can predict with certainty the value of a physical
quantity, then there exists an element of physical reality
corresponding to this physical quantity.} \vskip.7pc\noindent (L)
{\sl Principle of locality}.\quad{\it Let ${\mathcal R}_1$ and
${\mathcal R}_2$ be two space-like separated regions. The reality in
${\mathcal R}_2$ is unaffected by operations performed in ${\mathcal
R}_1$.} \vskip1pc\noindent In sections 3, 4, 5, we give equivalent
reformulations of the main non-locality theorems, which show that
what each of them proves is an inconsistency of quantum mechanics
with the following assumption involving quantum correlations.
\vskip0.7pc\noindent (EQC) {\sl Extension of quantum
correlations}.\quad {\it Let $A$ and $B$ be two observables whose
measurements require operations in regions ${\mathcal R}_1$ and
${\mathcal R}_2$, respectively, space-like separated from each
other. If quantum mechanics predicts correlations, in the state
$\psi$, between the outcomes of actually performed measurements of
$A$ and $B$, then every specimen $x$ of the physical system in the
state $\psi$ possesses objective values of $A$ and $B$ which satisfy
these correlations.} \vskip0.7pc\noindent The proved inconsistency
unavoidably leads to conclude that quantum mechanics violates the
principles of locality and reality if (EQC) could be inferred from
the principle of locality and reality, and, eventually, from quantum
theory. Such an inference can be established only if a {\it wide
interpretation} of the criterion of reality (R) is adopted,
expressed as law (wR) in section 2.
\par
The {\it strict} interpretation of (R), expressed as (sR) in
section 2, in not able to justify (EQC). Indeed, this last
interpretation leads to an extension (sEQC) of quantum
correlations which is strictly smaller than (EQC).
\par
In this work we explore the possibility of interpreting the
inconsistency proved by non-locality theorems as a failure of just
(EQC), without provoking conflicts between the principle (R,L) and
quantum mechanics. This program entails that interpretation (wR)
must be abandoned, because it implies (EQC); but (sR) must be
retained, because it is implied by (R,L). Furthermore, contrary to
(EQC), statement (sEQC) should not conflict with quantum mechanics,
because it is a consequence of (sR). In fact, we show that by
replacing (EQC) by (sEQC) in the main non-locality theorems, no
inconsistency between quantum theory and the principle of locality
and reality can be proved. Thus, the meaning of the present work is
that quantum mechanics can coexist with locality if the criterion of
reality is interpreted according to its strict sense.
\par
In section 2 we show how by strictly interpreting the criterion of
reality (R), only a weaker form (sEQC) of the extension law (EQC)
can be derived. Sections 3, 4, and 5 are devoted to show that the
non-locality theorems fail when the strict interpretation is
adopted, i.e. when (sEQC) replaces (EQC). We have proved such a
failure for the theorem \cite{6} proved by Greenberger, Horne,
Shimony, Zeilinger (GHSZ) in section 3, for the theorem of Hardy
\cite{4} in section 4 and for the classic Bell's theorem \cite{1}
in section 5. In fact, these three theorems follow three different
logical schemes, every non-locality theorem can be traced back to.

\section{Weak extension of quantum correlations.}
In this section we derive an extension of quantum correlations from
(R,L), as it can be inferred by a strict interpretation of the
criterion of reality (R). Our argument requires the formal
introduction of terms to suitably represent the concepts to be
handled. Given a quantum state vector $\psi$ of the Hilbert space
$\mathcal H$ which describes the physical system, let ${\mathcal
S}(\psi)$ be a {\it support} of $\psi$, i.e. a concrete set of
specimens of the physical systems whose quantum state is $\psi$. Let
$A$ be any 1-0 observable, i.e. an observable having only the
possible values 1 and 0, and hence represented by a projection
operator $\hat A$. In correspondence with $A$ we introduce the
following peculiar subsets of ${\mathcal S}(\psi)$. By ${\mathcal
A}$ we denote the set of the specimens of ${\mathcal S}(\psi)$ which
objectively possess a value of the observable $A$. By ${\mathcal
A}_1$ (resp., ${\mathcal A}_0$) we denote the set of specimens of
${\mathcal A}$ which possess the objective value 1 (resp., 0) of
$A$; hence we can assume that ${\mathcal A}_1\cup {\mathcal
A}_0={\mathcal A}$ holds. By ${\bf A}$ we denote the concrete set of
specimens of ${\mathcal S}(\psi)$ which actually undergo a
measurement of $A$. By ${\bf A}_1$ (resp., ${\bf A}_0$) we denote
the set of specimens of ${\bf A}$ for which the outcome 1 (resp., 0)
of $A$ has been obtained; hence we can assume that ${\bf A}_1\cup
{\bf A}_0={\bf A}$ holds. Moreover, we define the two mappings
$a:{\mathcal A}\to\{1,-1\}$ and ${\bf a}:{\bf A}\to\{1,-1\}$ as
follows.
$$a(x)=\left\{
\begin{array}{ll}\;\;\;1,\;x\in{\mathcal A}_1\, ,\\
-1, \; x\in{\mathcal A}_0\, ;\\
 \end{array}\right.
 \quad
{\bf a}(x)=\left\{
\begin{array}{ll}\;\,\;1,\;x\in{\bf A}_1\, ,\\
-1,\;x\in{\bf A}_0\, .\\
 \end{array}\right.
\eqno(1)
 $$
According to standard quantum theory the following statement holds
$${\bf A}_1\subseteq{\mathcal A}_1\hbox{ and }{\bf A}_0\subseteq{\mathcal A}_0,
\eqno(2.i)
$$
because the reality of the outcome of an actually performed
measurement cannot be denied. Moreover, two observables $A$ and
$B$ can be measured together if and only if the corresponding
operators commute with each other; therefore the following
statements hold.
$$
[\hat A, \hat B]\neq{\bf 0}\quad\hbox{implies}\quad {\bf
A}\cap{\bf B}=\emptyset\hbox{ for all }{\mathcal
S}(\psi).\eqno(2.ii)
$$
$$
\quad [\hat A, \hat B]={\bf 0}\quad\hbox{ implies }
\quad\forall\psi\;\;\exists{\mathcal S}(\psi)\hbox{
such that }{\bf A}\cap{\bf B}\neq\emptyset.\eqno(2.iii)
$$
\vskip.7pc Let $A$ and $B$ be two {\it separated} observables,
written $A\bowtie B$, i.e. observables whose measurements require
operations to be performed in space-like separated regions
${\mathcal R}_1$ and ${\mathcal R}_2$. Since (L) applies, the
following statement holds.
$$
A\bowtie B\quad\hbox{implies}\quad[\hat A,\hat B]={\bf 0},\;\hbox{
hence } {\mathcal S}(\psi) \hbox{ exists such that } {\bf
A}\cap{\bf B}\neq\emptyset. \eqno(3)
$$
\par
The principle of locality and reality (R,L) leads to further
implications in the case that the separated observables $A$ and
$B$ are {\it correlated}. Let us consider the case that the
correlation $A\to B$ holds in the quantum state $\psi$, which
means that whenever both $A$ and $B$ are measured, i.e. if
$x\in{\bf A}\cap{\bf B}$, then ${\bf a}(x)=1$ implies ${\bf
b}(x)=1$. Hence, the correlation $A\to B$ holds if and only if
${\bf A}_1\cap {\bf B}\subseteq {\bf B}_1$ or if and only if
$({a}(x)+1)({b}(x)-1)=0$ for all $x\in{\bf A}\cap {\bf B}$;
equivalently, $A\to B$ if and only if ${\bf B}_0\cap {\bf
A}\subseteq {\bf A}_0$. Now, if $A\bowtie B$ and  $A$ is measured
on $x\in{\bf A}$ obtaining ${\bf a}(x)=1$, then the principle of
locality (L) and the criterion of reality (R) imply $x\in{\mathcal
B}$ and $b(x)=1$. Therefore, ${\bf A}_1\subseteq {\mathcal B}_1$
and the correlation $(a(x)=1)\Rightarrow (b(x)=1)$ holds for all
$x\in{\bf A}_1$. The reasoning repeated by exchanging $A$ with $B$
leads to conclude that ${\bf B}_0\subseteq {\mathcal A}_0$ and
that the correlation $(a(x)=1)\Rightarrow (b(x)=1)$ holds for
every $x\in{\bf B}_0$. Thus the correlation extends to ${\bf
A}_1\cup{\bf B}_0$. Hence from (R,L) and quantum mechanics we
infer the following statement. \vskip.7pc\noindent (sEQC) {\sl
Weak extension of quantum correlations}.\quad{\it Let $A$ and $B$
be space-like separated 1-0 observables. If $A\to B$ then
$$\quad(a(x)+1)(b(x)-1)=0,\;\;\forall x\in({\bf
A}_1\cup{\bf B}_0)\cup({\bf A}\cap{\bf B}) .\eqno(4.i)
$$}
\vskip.7pc \noindent The quantum correlation $A\leftrightarrow B$,
i.e. $A\to B$ and $B\to A$, means that  the correlation
$(a(x)=1)\Leftrightarrow (b(x)=1)$ holds for all $x\in{\bf
A}\cap{\bf B}$. In this case, by (sEQC) we derive that
$(a(x)=1)\Leftrightarrow (b(x)=1)$ holds for all $x\in({\bf
A}_1\cup{\bf B}_0)\cup({\bf B}_1\cup{\bf A}_0)\cup({\bf A}\cap{\bf
B})={\bf A}\cup{\bf B}$. Hence, (sEQC) incorporates the following
extension of quantum correlations.
$$A\bowtie B\hbox{ and }A\leftrightarrow B\quad\hbox{imply}
\quad a(x)=b(x),\;\; \forall x\in{\bf A}\cup{\bf B}. \eqno(4.ii)
$$
\par
We remark that in deriving (sEQC) we have applied the {\it strict}
interpretation (sR) of the criterion of reality, according to
which if $A\bowtie B$ and $A\to B$ we can predict with certainty
the value of an eventual measurement of $B$ (resp., $A$) only once
a measurement of $A$ with concrete outcome ${\bf a}(x)=1$ (resp.,
$B$ with concrete outcome ${\bf b}(x)=0$) is performed. If
$x\notin{\bf A}_1$ (resp., $x\notin{\bf B}_0$)  no prediction of
the value of $B$ (resp., $A$) can be made by a strict application
of (R). \vskip.7pc The larger extension stated by (EQC) can be
derived from (R,L) only if a wider interpretation (wR) is adopted,
according to which for ascribing reality to $B$ it is sufficient
the ``possibility'' of performing a measurement whose outcome
would allow for the prediction, with certainty, of the outcome of
an eventual measurement of $B$. Note 10 in \cite{6} highlights the
importance of this twofold possibility in interpreting the
criterion of reality.
%%%%%%%%%%%%%%%%%%%%%%%%%%%%%%%%%%%%%%%%%%%%%%%%%%%%%%%%%%%%%%%%%%%%%%%%%%%%%%%%%%%%%%%%%%%%%
\section{GHSZ theorem does not extend to (sEQC)}
In this section we show that the argument of GHSZ cannot be used
for proving inconsistency between quantum mechanics and statement
(sEQC). In so doing, we first reformulate GHSZ proof to make clear
the role of law (EQC). \vskip.7pc The theorem of GHSZ makes use of
seven 1-0 observables of a particular quantum system, separated
into four classes
$$
\omega_A=\{A^\alpha,A^\beta\},\;\omega_B=\{B\},
\;\omega_C=\{C^\alpha,C^\beta\},\;\omega_D=\{D^\alpha,D^\beta\}.
$$
These observables have been singled out by GHSZ in such a way that
\vskip.5pc\noindent (5.i)\quad two observables in two different
classes are separated from each other. \vskip.5pc\noindent (5.ii)
\quad $[\hat{A^\alpha},\hat{A^\beta}]\neq{\bf 0}$,
$[\hat{C^\alpha},\hat{C^\beta}]\neq{\bf 0}$,
$[\hat{D^\alpha},\hat{D^\beta}]\neq{\bf 0}$. \vskip.5pc\noindent
The state vectors $\psi$ is chosen so that the following
correlations between actually measured outcomes hold, according to
quantum mechanics. {\setlength\arraycolsep{2pt}
$$
\begin{array}{llll}
\textrm{   i)}\quad &{\bf a}^\alpha(x){\bf b}(x)&=-{\bf
c}^\alpha(x){\bf d}^\alpha(x)\quad
&\forall x\in({\bf A}^\alpha\cap{\bf B})\cap({\bf C}^\alpha\cap{\bf D}^\alpha)\equiv{\bf X},\\
\textrm{  ii)}\quad &{\bf a}^\beta(y){\bf b}(y)&=-{\bf
c}^\beta(y){\bf d}^\alpha(y)\quad
&\forall y\in({\bf A}^\beta\cap{\bf B})\cap({\bf C}^\beta\cap{\bf D}^\alpha)\equiv{\bf Y},\\
\textrm{ iii)}\quad &{\bf a}^\beta(z){\bf b}(z)&=-{\bf
c}^\alpha(z){\bf d}^\beta(z)\quad
&\forall z\in({\bf A}^\beta\cap{\bf B})\cap({\bf C}^\alpha\cap{\bf D}^\beta)\equiv{\bf Z},\\
\textrm{  iv)}\quad &{\bf a}^\alpha(t){\bf b}(t)&={\bf
c}^\beta(t){\bf d}^\beta(t)\quad
&\forall t\in({\bf A}^\alpha\cap{\bf B})\cap({\bf C}^\beta\cap{\bf D}^\beta)\equiv{\bf T}.\\
\end{array}
\eqno(6)
$$}
In terms of 1-0 observables, equations (6.i), (6.ii), (6.iii),
(6.iv) express the quantum correlations $A^\alpha \ast
B\leftrightarrow 1-C^\alpha\ast D^\alpha$, $A^\beta\ast
B\leftrightarrow 1-C^\beta\ast D^\alpha$, $A^\beta\ast
B\leftrightarrow 1-C^\alpha\ast D^\beta$, $A^\alpha\ast B
\leftrightarrow C^\beta\ast D^\beta$, respectively, where we have
put $A\ast B=1-(A-B)^2$.
\par
If (EQC) holds, then correlations (6) must be
extended to the following correlations between objective values.
{\setlength\arraycolsep{2pt}
$$
\left.
\begin{array}{llll}
\textrm{   i)}\quad &{a}^\alpha(x){b}(x)&=-{c}^\alpha(x){d}^\alpha(x),\\
%{bf A}^\alpha\cap{\bf B}\cap{\bf C}^\alpha\cap{\bf D}^\alpha,\\
\textrm{  ii)}\quad &{a}^\beta(x){b}(x)&=-{c}^\beta(x){d}^\alpha(x),\\
%\forall y\in{\bf A}^\beta\cap{\bf B}\cap{\bf C}^\beta\cap{\bf D}^\alpha,\\
\textrm{ iii)}\quad &{a}^\beta(x){b}(x)&=-{c}^\alpha(x){d}^\beta(x),\\
%\forall z\in{\bf A}^\beta\cap{\bf B}\cap{\bf C}^\alpha\cap{\bf D}^\beta,\\
\textrm{  iv)}\quad &{a}^\alpha(x){b}(x)&={c}^\beta(x){d}^\beta(x),\\
%\forall t\in{\bf A}^\alpha\cap{\bf B}\cap{\bf C}^\beta\cap{\bf D}^\beta.\\
\end{array}\right\}
\quad\forall x\in{\mathcal S}(\psi). \eqno(7)
$$}
GHSZ prove that the correlations (7) are inconsistent because
(i)-(iv) in (7) hold for a same $x\in{\mathcal
S}(\psi)\neq\emptyset$. Indeed, from (7.i) and (7.iv) we get
$$
{c}^\alpha(x){d}^\alpha(x)=-{c}^\beta(x){d}^\beta(x). \eqno(8)
$$
On the other hand, from (7.ii) and (7.iii) the equality
${c}^\alpha(x){d}^\beta(x)={c}^\beta(x){d}^\alpha(x)$ follows,
which is equivalent, since
${d}^\beta(x){d}^\beta(x)={d}^\alpha(x){d}^\alpha(x)=1$, to
$$
{ c}^\alpha(x){d}^\alpha(x)={c}^\beta(x){d}^\beta(x) \eqno(9)
$$
which contradicts (8). \vskip.7pc Now we prove that this proof of
inconsistency does not work if we replace (EQC) by (sEQC). To this
end, it is worth to remark that the contradiction between (8) and
(9) cannot be derived from (6) alone, without the extension to (7)
implied by (EQC), because (6.i)-(6.iv) cannot hold simultaneously
for a same $x=y=z=t$; indeed,
$[\hat{A^\alpha},\hat{A^\beta}]\neq{\bf 0}$ by (5.ii); hence,
according to quantum theory, ${\bf A}^\alpha\cap{\bf
A}^\beta=\emptyset$ by (2.ii), and therefore ${\bf X}\cap{\bf
Y}\cap{\bf Z}\cap{\bf T}=\emptyset$.
\par
The extension of correlations (6) validated by (sEQC) in this case
are obtained by applying (4.ii), i.e. {\setlength\arraycolsep{2pt}
$$
\begin{array}{llll}
\textrm{   i)}\quad &{ a}^\alpha(x){ b}(x)=-{ c}^\alpha(x){
d}^\alpha(x)\quad
&\forall x\in({\bf A}^\alpha\cap{\bf B})\cup({\bf C}^\alpha\cap{\bf D}^\alpha)\equiv{\tilde{\bf X}},\\
\textrm{  ii)}\quad &{   a}^\beta(y){   b}(y)=-{   c}^\beta(y){
d}^\alpha(y)\quad
&\forall y\in({\bf A}^\beta\cap{\bf B})\cup({\bf C}^\beta\cap{\bf D}^\alpha)\equiv{\tilde{\bf Y}},\\
\textrm{ iii)}\quad &{   a}^\beta(z){   b}(z)=-{ c}^\alpha(z){
d}^\beta(z)\quad
&\forall z\in({\bf A}^\beta\cap{\bf B})\cup({\bf C}^\alpha\cap{\bf D}^\beta)\equiv{\tilde{\bf Z}},\\
\textrm{  iv)}\quad &{   a}^\alpha(t){   b}(t)={   c}^\beta(t){
d}^\beta(t)\quad
&\forall t\in({\bf A}^\alpha\cap{\bf B})\cup({\bf C}^\beta\cap{\bf D}^\beta)\equiv{\tilde{\bf T}}.\\
\end{array}
\eqno(10)
$$
In order that the GHSZ argument -- which leads to the contradiction
from (7) to (9) through (8)-- can be successfully repeated from
(10), the first step requires that (10.i) and (10.iv) should hold
for the same specimen $x_0$; therefore the condition ${\tilde{\bf
X}}\cap{\tilde{\bf T}}\neq\emptyset$ should hold; the second step
requires that also (10.ii) and (10.iii) should hold for such a
specimen. Thus, the condition
$${\tilde{\bf X}}\cap{\tilde{\bf
Y}}\cap{\tilde{\bf Z}}\cap{\tilde{\bf T}}\neq\emptyset\eqno(11)$$
should be satisfied. Now, from (5.ii) and (2.ii) we derive
$$\emptyset=
({\bf A}^\alpha\cap{\bf B})\cap({\bf A}^\beta\cap{\bf B})= ({\bf
C}^\alpha\cap{\bf D}^\alpha)\cap({\bf C}^\beta\cap{\bf D}^\alpha)=
$$ $$=({\bf C}^\alpha\cap{\bf D}^\alpha)\cap({\bf C}^\alpha\cap{\bf
D}^\beta)= ({\bf C}^\alpha\cap{\bf D}^\alpha)\cap({\bf
C}^\beta\cap{\bf D}^\beta)= \eqno(12)$$ $$= ({\bf C}^\beta\cap{\bf
D}^\alpha)\cap({\bf C}^\alpha\cap{\bf D}^\beta)= ({\bf
C}^\beta\cap{\bf D}^\alpha)\cap({\bf C}^\beta\cap{\bf D}^\beta)=$$
$$
=({\bf C}^\alpha\cap{\bf D}^\beta)\cap({\bf C}^\beta\cap{\bf
D}^\beta).
$$
To obtain the set ${\tilde{\bf X}}\cap{\tilde{\bf
Y}}\cap{\tilde{\bf Z}}\cap{\tilde{\bf T}}$ the distributive law
for $\cap$ and $\cup$ of elementary set theory can be applied; in
so doing, (12) imply
$${\tilde{\bf X}}\cap{\tilde{\bf Y}}\cap{\tilde{\bf
Z}}\cap{\tilde{\bf T}}=\emptyset,
$$
which denies condition (11) necessary to prove the inconsistency.
Thus, GHSZ proof cannot be extended to prove inconsistency between
quantum mechanics and the strict interpretation of the principle
of locality and reality.

%%%%%%%%%%%%%%%%%%%%%%%%%%%%%
%%%%%%%%%%%%%%%%%%%%%%%%%%%%%
\section{Hardy's theorem}
The scheme of the theorem of Hardy involves four 1-0 observables
$A^\alpha$, $B^\alpha$, $A^\beta$, $B^\beta$, chosen in such a way
that \vskip.5pc\noindent (13.i)\quad $A^\alpha\bowtie B^\alpha$,
$A^\alpha\bowtie B^\beta$, $A^\beta\bowtie B^\alpha$
$A^\beta\bowtie B^\beta$;
\par\noindent
(13.ii)\quad $[\hat{A^\alpha},\hat{A^\beta}]\neq{\bf 0}$ and
$[\hat{B^\alpha},\hat{B^\beta}]\neq{\bf 0}$. \vskip.5pc\noindent The
state vector $\psi$ is chosen so that according to quantum theory
the correlations $A^\alpha\to B^\alpha$, $B^\alpha\to A^\beta$,
$A^\beta\to B^\beta$ hold, which can be equivalently expressed as
follows. {\setlength\arraycolsep{2pt}
$$
\begin{array}{lll}
\textrm{   i)}\quad &(a^\alpha(x)+1)(b^\alpha(x)-1)&=0,\quad \forall
x\in{\bf A}^\alpha\cap{\bf B}^\alpha
\\
\textrm{  ii)}\quad &(b^\alpha(y)+1)(a^\beta(y)-1)&=0,\quad \forall
y\in{\bf A}^\beta\cap{\bf B}^\alpha
\\
\textrm{ iii)}\quad &(a^\beta(z)+1)(b^\beta(z)-1)&=0,\quad \forall
z\in{\bf A}^\beta\cap{\bf B}^\beta.
\\
\end{array}
\eqno(14)
$$}
A further constraint satisfied by the choice of $\psi$ in Hardy's theorem
is the
following statistical prediction of quantum mechanics,
$$
\langle\psi\mid \hat{A^\alpha}({\bf
1}-\hat{B^\beta})\psi\rangle\neq 0. \eqno(15.i)
$$
which means that there is a non-vanishing probability of obtaining
$(1,0)$ as pair of outcomes of a measurement of $A^\alpha$ and
$B^\beta$. Hence, a support ${\mathcal S}(\psi)$ exists such that
${\bf A}^\alpha_1\cap{\bf B}^\beta_0\neq\emptyset$, i.e.
$$
{\mathcal S}(\psi)\hbox{ and }x_0\in{\mathcal S}(\psi)\hbox{ exist}
\quad\hbox{such that}\quad a^\alpha(x_0)=1\hbox{ and
}b^\beta(x_0)=-1.\eqno(15.ii)
$$
If (EQC) is assumed to hold, then from correlations (14) we infer
that {\setlength\arraycolsep{2pt}
$$
\begin{array}{lll}
\textrm{   i)}\quad &(a^\alpha(x)+1)(b^\alpha(x)-1)&=0,
\\
\textrm{  ii)}\quad &(b^\alpha(x)+1)(a^\beta(x)-1)&=0,
\\
\textrm{ iii)}\quad &(a^\beta(x)+1)(b^\beta(x)-1)&=0
\end{array}
\eqno(16)
$$}
are satisfied for any
$x\in{\mathcal S}(\psi)$, for every support ${\mathcal
S}(\psi)$.
Now,
if a specimen $x$ satisfies (16.i,ii,iii), then by using elementary algebra
we imply that
$$
(a^\alpha(x)+1)(b^\beta(x)-1)=0 \eqno(17) $$ holds for such $x$.
Therefore (17) holds for every $x\in{\mathcal S}(\psi)$, for every
support ${\mathcal S}(\psi)$. Thus, (15.ii) turns out to be
contradicted, because $(a^\alpha(x_0)+1)(b^\beta(x_0)-1)=-4$.
\vskip.7pc Now we show that no contradiction arises if we replace
(EQC) by (sEQC). The extension of correlations (14) obtained by
applying (sEQC) is expressed by {\setlength\arraycolsep{2pt}
$$
\begin{array}{llll}
\textrm{   i)}\quad &(a^\alpha(x)+1)(b^\alpha(x)-1)&=0,\quad
&\forall x\in{\bf A}_1^\alpha\cup{\bf B}_0^\alpha\cup({\bf
A}^\alpha\cap{\bf B}^\alpha)={\bf X}
\\
\textrm{  ii)}\quad &(b^\alpha(y)+1)(a^\beta(y)-1)&=0,\quad &\forall
y\in{\bf A}_1^\beta\cup{\bf B}_0^\alpha\cup({\bf A}^\beta\cap{\bf
B}^\alpha)={\bf Y}
\\
\textrm{ iii)}\quad &(a^\beta(z)+1)(b^\beta(z)-1)&=0,\quad &\forall
z\in{\bf A}_1^\beta\cup{\bf B}_0^\beta\cup({\bf A}^\beta\cap{\bf
B}^\beta)={\bf Z}.
\\
\end{array}
\eqno(18)
$$}
These extensions no longer imply (17). Indeed, equation
$(a^\alpha(x)-1)(b^\beta(x)-1)=0$ can be derived from (18) if all
three equations therein hold for the same specimen $x$, i.e. if
$x\in{\bf X}\cap{\bf Y}\cap{\bf Z}$. But this last set is empty;
indeed $({\bf A}^\alpha\cup{\bf B}^\alpha)\cap({\bf A}^\beta\cup{\bf
B}^\alpha)\cap({\bf A}^\beta\cup{\bf B}^\beta)=\emptyset$ follows
from (13.ii) and (2.ii); on the other hand, from the definition of
${\bf X}$, $\bf Y$ and $\bf Z$ in (18) we have ${\bf X}\cap{\bf
Y}\cap{\bf Z}\subseteq({\bf A}^\alpha\cup{\bf B}^\alpha)\cap({\bf
A}^\beta\cup{\bf B}^\alpha)\cap({\bf A}^\beta\cup{\bf B}^\beta)$,
and thus ${\bf X}\cap{\bf Y}\cap{\bf Z}=\emptyset$.

%%%%%%%%%%%%%%%%%%%%%%%%%%%%%%%%%%%%%%%%%%%%%%%%%%%%%%%%%%%%%%%%%%%%%%%%%%%%%
\section{\bf Bell's theorem}
Six 1-0 observables $A^\alpha$, $A^\beta$, $A^\gamma$, $B^\alpha$,
$B^\beta$, $B^\gamma$ are involved in Bell's theorem, which satisfy
the following conditions \vskip.5pc\noindent (19.i)\quad Each
$A$-observable is space-like separated from every $B$-observable;
\vskip.5pc\noindent (19.ii)\quad
$[\hat{A^\lambda},\hat{A^\mu}]\neq{\bf 0}$ and
$[\hat{B^\lambda},\hat{B^\mu}]\neq{\bf 0}$ if
$\lambda\neq\mu$,\hbox{ where }
$\lambda,\mu\in\{\alpha,\beta,\gamma\}$. \vskip.5pc\noindent The
state vector $\psi$ is singled out so that quantum correlations
$A^\beta\leftrightarrow 1- B^\beta$ and $A^\gamma\leftrightarrow
1-B^\gamma$ hold, i.e.
$$ \hbox{(i)}\quad a^\beta(x)=-b^\beta(x)\qquad \hbox{and\qquad(ii)\quad
}a^\gamma(x)=-b^\gamma(x), \eqno(20)
$$
where (20.i) holds for all
$x\in{\bf
A}^\beta\cap{\bf B}^\beta$ and (20.ii) holds for all
$x\in{\bf A}^\gamma\cap{\bf
B}^\gamma$.
Following Bell's proof, if
$Y=\{x_1,x_2, \ldots,x_N\}$ is any finite set of specimens of
the physical system such that both (20.i) and (20.ii) hold for every $x_k\in Y$,
then the following (Bell's) inequality
$$
\left\vert\overline{a^\alpha b^\beta}-\overline{a^\alpha
b^\gamma}\right\vert\leq 1-\overline{a^\beta
b^\gamma}
\eqno(21)
$$
can be derived for the mean values $\overline{a^\alpha b^\beta}$,
$\overline{a^\alpha b^\gamma}$, $\overline{a^\beta b^\gamma}$
(for instance, $\overline{a^\alpha b^\beta} = [\sum_{x_k\in
Y}a^\alpha(x_k) b^\beta(x_k)]/N$, and so on), all computed on the
same sample $Y$.
Quantum mechanics, by itself, does not conflict with Bell's inequality (21),
because according to quantum theory the correlations (20.i) and (20.ii)
hold together only if $x\in{\bf A}^\beta\cap{\bf B}^\beta\cap{\bf A}^\gamma\cap{\bf
B}^\gamma=\emptyset$.
\par
But if (EQC) is assumed to hold, then (20) extends to
$$ \hbox{(i)}\quad a^\beta(x)=-b^\beta(x),\;\forall x\in{\mathcal S}(\psi);\qquad
\hbox{(ii)}\quad a^\gamma(x)=-b^\gamma(x),\;\forall x\in {\mathcal S}(\psi).
\eqno(22)
$$
Therefore, (EQC) makes valid Bell's inequality (21) whenever the
involved mean values are evaluated for the ``objective'' values on
{\it any} finite sample $Y\subseteq{\mathcal S}(\psi)$. Now, quantum
theory cannot predict the three mean values in (21) evaluated on a
same sample $Y$, because they refer to the three non-commuting (by
19.ii) observables $A^\alpha\ast B^\beta$, $A^\alpha\ast B^\gamma$,
$A^\beta \ast B^\gamma$. Quantum mechanics can predict the mean
values $\overline{a^\alpha b^\beta}$, $\overline{a^\alpha
b^\gamma}$, $\overline{a^\beta b^\gamma}$ evaluated on {\it
different} samples $Y_1\subseteq{\bf A}^\alpha\cap{\bf B}^\beta$,
$Y_2\subseteq{\bf A}^\alpha\cap{\bf B}^\gamma$, $Y_3\subseteq{\bf
A}^\beta\cap{\bf B}^\gamma$, each of them contained in the domain of
validity of (21); these mean values agree, according to quantum
theory, with the quantum expectation values
$\langle\psi\mid\hat{A^\alpha}\ast\hat{B^\beta}\mid\psi\rangle$,
$\langle\psi\mid\hat{A^\alpha}\ast\hat{B^\gamma}\mid\psi\rangle$,
$\langle\psi\mid\hat{A^\beta}\ast\hat{B^\gamma}\mid\psi\rangle$. If
the mean values are replaced by these expectation values, a
violation of Bell's inequality (21) is found\footnote{In order that
the mean values in (21) can be replaced by the quantum expectation
values, the following further hypothesis has been assumed in Bell's
type theorems. \vskip.5pc\noindent {\sl Fair sampling
assumption.}\quad{\it The sample of physical systems which actually
undergo a measurement fairly represent the entire population
${\mathcal S}(\psi)$.} \vskip.5pc\noindent The validity of the fair
sampling assumption has been submitted to deep investigations, as
for instance in \cite{8,9} and references therein, which show that
it can be seriously questioned without violating physical principles
or statistical regularity. This assumption is not necessary in the
theorems of GHSZ and Hardy which, for this reason, are more
effective in showing inconsistency between quantum mechanics and
(EQC).}. \vskip.7pc Now we show that if (EQC) is replaced by (sEQC)
then the domain of validity of (21) becomes smaller than ${\mathcal
S}(\psi)$, so that the predictions of quantum theory about the mean
values in (21) cannot longer apply, because they refer to samples
$Z$ which are outside of the domain of validity of (21).
\par
By using (4.ii), we imply that (20.i) holds for all $x\in{\bf
A}^\beta\cup{\bf B}^\beta$, whereas (20.ii) is valid for all
$x\in{\bf A}^\gamma\cup{\bf B}^\gamma$; therefore both (20.i) and
(20.ii) hold for all $x\in {\bf X}= ({\bf A}^\beta\cup{\bf
B}^\beta)\cap({\bf A}^\gamma\cup{\bf B}^\gamma)$. Since $[\hat{
A}^\beta,\hat{ A}^\gamma]\neq{\bf 0}$ and $[\hat{ B}^\beta,\hat{
B}^\gamma]\neq{\bf 0}$ we have ${\bf X}=({\bf A}^\beta\cup{\bf
B}^\beta)\cap({\bf A}^\gamma\cup{\bf B}^\gamma)=({\bf
A}^\beta\cap{\bf A}^\gamma)\cup({\bf A}^\beta\cap{\bf
B}^\gamma)\cup({\bf B}^\beta\cap{\bf A}^\gamma)\cup(\hat{\bf
B}^\beta\cap\hat{\bf B}^\gamma)= ({\bf A}^\beta\cap{\bf
B}^\gamma)\cup({\bf B}^\beta\cap{\bf A}^\gamma)$ by (2.ii). On the
other hand by (19.i), (3) and (2.iii) we infer that a support
${\mathcal S}(\psi)$ exists such that $({\bf A}^\beta\cap{\bf
B}^\gamma)\neq\emptyset$ (or $({\bf A}^\gamma\cap {\bf
B}^\beta)\neq\emptyset$), and hence ${\bf X}\neq\emptyset$.
Therefore, a result of the substitution of (EQC) with (sEQC) is that
Bell inequality (21) holds only for samples ${\bf Y}\subseteq{\bf
X}=({\bf A}^\beta\cap{\bf B}^\gamma)\cup({\bf B}^\beta\cap{\bf
A}^\gamma)$. \par Such a limited validity {\it does not violates
quantum mechanics}. Indeed, the quantum mechanical prediction
$\langle \psi\mid\hat{A^\alpha}\ast\hat{B^\beta}\mid \psi\rangle$
for $\overline{a^\alpha b^\beta}$ in (21) holds for samples $Z_1$
such that at least $Z_1\subseteq{\bf A}^\alpha$ is satisfied. But
since $[\hat{ A}^\alpha,\hat{ A}^\beta]\neq{\bf 0}$ and $[\hat{
A}^\alpha,\hat{ A}^\gamma]\neq{\bf 0}$, by (2.ii) we have
$$
{\bf A}^\alpha\cap{\bf X}=({\bf A}^\alpha\cap{\bf A}^\beta\cap{\bf B}^\gamma)\cup
({\bf A}^\alpha\cap{\bf A}^\gamma\cap{\bf B}^\beta)=\emptyset.\eqno(23)
$$
Hence the quantum mechanical prediction for
$\overline{a^\alpha b^\beta}$ refer to samples for which the limited
Bell's inequality does not hold.
\par
To conclude, (sEQC) bounds the validity of Bell inequality  in
such a way that \vskip.5pc\noindent (i)\quad it holds for samples
which make Bell inequality neither verifiable by experiment, nor
comparable with the statistical predictions of quantum theory;
\vskip.5pc\noindent (ii)\quad it is consistent with the principle
of reality and locality and with the predictions of quantum
theory.

\end{document}